\documentclass[a4paper]{ESLAB2005}
\usepackage{epsfig}
\usepackage{graphicx}

\usepackage{dcolumn}

\begin{document}

\title{Fundamental Physics with the Laser Astrometric Test Of Relativity}

\author{
S.\,G.~Turyshev,\inst{1}
H.~Dittus,\inst{2} C.~L\"ammerzahl,\inst{2} S.~Theil,\inst{2}
W.~Ertmer,\inst{3} E.~Rasel,\inst{3}
R.~Foerstner,\inst{4} U.~Johann,\inst{4}
S.~Klioner,\inst{5} M.~Soffel,\inst{5} 
B.~Dachwald,\inst{6} W.~Seboldt,\inst{6}
V.~Perlick,\inst{7}
M.\,C.\,W.~Sandford,\inst{8} R.~Bingham,\inst{8} B.~Kent,\inst{8}
T.\,J.~Sumner,\inst{9}
O.~Bertolami,\inst{10} J.~P\'aramos,\inst{10}
B.~Christophe,\inst{11} B.~Foulon,\inst{11} P.~Touboul,\inst{11}
P.~Bouyer,\inst{12} 
T.~Damour,\inst{13}
C.~Salomon,\inst{14} S.~Reynaud,\inst{14}
A.~Brillet,\inst{15} F.~Bondu,\inst{15} J.-F.~Mangin,\inst{15} 
E.~Samain,\inst{15} 
C.~Erd,\inst{16} J.\,C.~Grenouilleau,\inst{16} D.~Izzo,\inst{16} A.~Rathke,\inst{16}
S.\,W.~Asmar,\inst{1} M.\,Colavita,\inst{1} Y.~G\"ursel,\inst{1} H.~Hemmati,\inst{1} M.~Shao,\inst{1} 
J.\,G.~Williams,\inst{1}
K.\,L.~Nordtvedt,~Jr.,\inst{17} 
J.~Degnan,\inst{18}   
J.\,E.~Plowman,\inst{19}  R.~Hellings,\inst{19} 
\and  
T.\,W.~Murphy, Jr.\inst{20}  
} 
\institute{
Jet Propulsion Laboratory, 
California Institute of Technology, 
4800 Oak Grove Drive, Pasadena, CA 91109, USA
\and
Centre of Applied Space Technology \& Microgravity (ZARM), 
University of Bremen, Am Fallturm, 
28359 Bremen, 
Germany
\and 
Institute for Quantum Optics, University of Hannover
Welfengarten 1,
30167 Hannover,
Germany 
\and 
Department of Science Programs,  
Earth Observation and Science,
Astrium GmbH, 
88039 Friedrichshafen,
Germany
\and 
Lorhmann Observatory, 
Dresden Technical University,
Mommsenstrasse, 13,
01062 Dresden, 
Germany
\and
German Aerospace Center (DLR), 
Institute of Space Simulation,
Linder Hoehe, D-51170 K\"oln, 
Germany
\and
Technical University of Berlin,
Institute for Theoretical Physics,
Hardenbergstrasse 36,
10623 Berlin,
Germany
\and 
Space Engineering and Technology Division,
Rutherford Appleton Laboratory,
Chilton,
Oxfordshire OX11 0QX,
UK
\and 
Imperial College, 
The Blackett Laboratory, 
Prince Consort Road, London SW7 2BZ, 
London, UK
\and 
Instituto Superior T\'ecnico, Departamento de F\'isica,  
Av. Rovisco Pais, 1, 1049-001 Lisboa, Portugal
\and 
Physics and Instrumentation Department, 
ONERA,
BP72,
29 ave de la division Leclerc,
92322 Chatillon,
France
\and 
Laboratoire Charles Fabry de l'Institut d'Optique,
Bat 503,
Centre Scientifique,
91403 ORSAY CEDEX,
France
\and
Institut des Hautes Etudes Scientifiques, 
35, route de Chartres, 
F-91440 Bures-sur-Yvette, France
\and
Laboratoire Kastler Brossel,
Universit\'e Pierre et Marie Curie,
Campus Jussieu case 74,
75252 Paris,
France
\and 
Observatoire de la C\^ote d'Azur, 
CERGA, Av. N. Copernic, 
F-06130 Grasse, France
\and
ESA Advanced Concepts Team,
ESTEC (SER-A)
Keplerlaan 1,
2201 AZ Noordwijk ZH,
The Netherlands	
\and 
Northwest Analysis, 118 Sourdough Ridge Road, Bozeman, MT 59715 USA
\and 
Sigma Space Corporation,
4801 Forbes Blvd.,
Lanham, MD 20706
USA 
\and
Physics Department,
Montana State University,
Bozeman, MT 59717 USA
\and
Physics Department, University of California, San Diego, CASS-0424, 9500 Gilman Dr., La Jolla, CA 92093 USA
}

\maketitle 

\begin{abstract}
\vskip -0pt
The Laser Astrometric Test Of Relativity (LATOR) is a joint European-U.S.  Michelson-Morley-type experiment designed to test the  metric nature of gravitation -- a fundamental postulate of Einstein's theory of general relativity.  By using a combination of independent time-series of highly accurate gravitational deflection of light in the immediate proximity to the Sun, along with measurements of the Shapiro time delay on interplanetary scales (to a precision respectively better than $10^{-13}$ radians and 1 cm), LATOR will significantly improve our knowledge of relativistic gravity.  
The primary mission objective is to i) measure the key post-Newtonian Eddington parameter $\gamma$ with accuracy of a part in 10$^9$.  $(1-\gamma)$ is a direct measure for presence of a new interaction in gravitational theory, and, in its search, LATOR goes a factor 30,000 beyond the present best result, Cassini's 2003 test.  Other mission objectives include: ii) first measurement of gravity's non-linear effects on light to $\sim$0.01\% accuracy; including both the traditional Eddington $\beta$  parameter and also the spatial metric's 2nd order potential contribution (never measured before);  iii) direct measurement of the solar quadrupole moment $J_2$ (currently unavailable) to accuracy of a part in 200 of its expected size; iv) direct measurement of the ``frame-dragging'' effect on light due to the Sun's rotational gravitomagnetic field, to 1\% accuracy. LATOR's primary measurement pushes to unprecedented accuracy the search for cosmologically relevant scalar-tensor theories of gravity by looking for a remnant scalar field in today's solar system. The key element of LATOR is a geometric redundancy provided by the laser ranging and long-baseline optical interferometry.  LATOR is envisaged as a partnership between European and US institutions and with clear areas of responsibility between the space agencies: NASA provides the deep space mission components, while optical infrastructure on the ISS is an ESA contribution. We discuss the mission and optical designs  of this proposed experiment. 

\vskip -8pt 
\end{abstract}

\section{Introduction}
\label{sec:intro}

After almost ninety years since general relativity was born, Einstein's general theory of relativity (GR) has survived every test. Such longevity, of course, does not mean that this theory is absolutely correct, but it serves to motivate more accurate tests to determine the level of accuracy at which it is violated. GR began its empirical success in 1915 by explaining the anomalous perihelion precession of Mercury's orbit, using no adjustable theoretical parameters.  Shortly thereafter, Eddington's 1919 observations of star lines-of-sight during a solar eclipse confirmed the doubling of the deflection angles predicted by GR, as compared to Newtonian-like and Equivalence Principle arguments.  This test made  general relativity an instant success. 

From these beginnings, the general theory of relativity has been verified at ever higher accuracy. Thus, microwave ranging to the Viking Lander on Mars yielded a $\sim$0.2\% accuracy in the tests of GR (\cite{viking_shapiro1,viking_reasen}). Spacecraft and planetary radar observations reached an accuracy of $\sim$0.15\% (\cite{anderson02}).  The astrometric observations of quasars on the solar background performed with Very-Long Baseline Interferometry (VLBI) improved the accuracy of the tests of GR to $\sim$0.045\% (\cite{Shapiro_SS_etal_2004}). Lunar laser ranging,  a continuing legacy of the Apollo program, provided $\sim$0.011\% verification of GR via precision measurements of the lunar orbit (\cite{Ken_LLR_PPNprobe03,LLR_beta_2004}). Finally, the recent experiments with the Cassini spacecraft improved the accuracy of the tests to $\sim$0.0023\% (\cite{cassini_ber}). As a result, general relativity became the standard theory of gravity when astrometry and spacecraft navigation are concerned. 

However, the tensor-scalar theories of gravity, where the usual general relativity tensor field coexists with one or several long-range scalar fields, are believed to be the most promising extension of the theoretical foundation of modern gravitational theory. The superstring, many-dimensional Kaluza-Klein, and inflationary cosmology theories have revived interest in the so-called `dilaton fields', i.e. neutral scalar fields whose background values determine the strength of the coupling constants in the effective four-dimensional theory. The importance of such theories is that they provide a possible route to the quantization of gravity and the unification of physical laws. 

Recent theoretical findings suggest that the present agreement between GR and experiment might be naturally compatible with the existence of a scalar contribution to gravity. In particular, \cite*{Damour_Nordtvedt_1993a} (see also \cite*{DamourPolyakov94} for non-metric versions of this mechanism together with \cite*{DPV02} for the recent summary of a dilaton-runaway scenario) have found that a scalar-tensor theory of gravity may contain a `built-in' cosmological attractor mechanism towards GR.  These scenarios assume that the scalar coupling parameter $\frac{1}{2}(1-\gamma)$ was of order one in the early universe (say, before inflation), and show that it then evolves to be close to, but not exactly equal to, zero at the present time. Under some assumptions (see e.g. \cite*{Damour_Nordtvedt_1993a,DPV02}) one can even estimate what is the likely order of magnitude of the left-over coupling strength at present time, with results in the range from $10^{-5}$ to $5\times10^{-8}$ for $(1-\gamma)$, i.e. for observable post-Newtonian deviations from general relativity predictions. This would require measurement of the effects of the next post-Newtonian order ($\propto G^2$) of light deflection resulting from gravity's intrinsic non-linearity. An ability to measure the first order light deflection term at the accuracy comparable with the effects of the second order is of the utmost importance for gravitational theory and a major challenge for the 21st century fundamental physics. 

Another attractive theoretical possibility to extend GR involves a putative breaking of Lorentz invariance through a nonvanishing vacuum expectation value of a vector field (\cite{Kostelecky_04}). This is a quite realistic scenario, for instance, in the context of string field theory. Solutions and potential observational implications have recently been examined (\cite{BP-1}).

\begin{figure*}[t!]
 \begin{center}
\noindent    
\psfig{figure=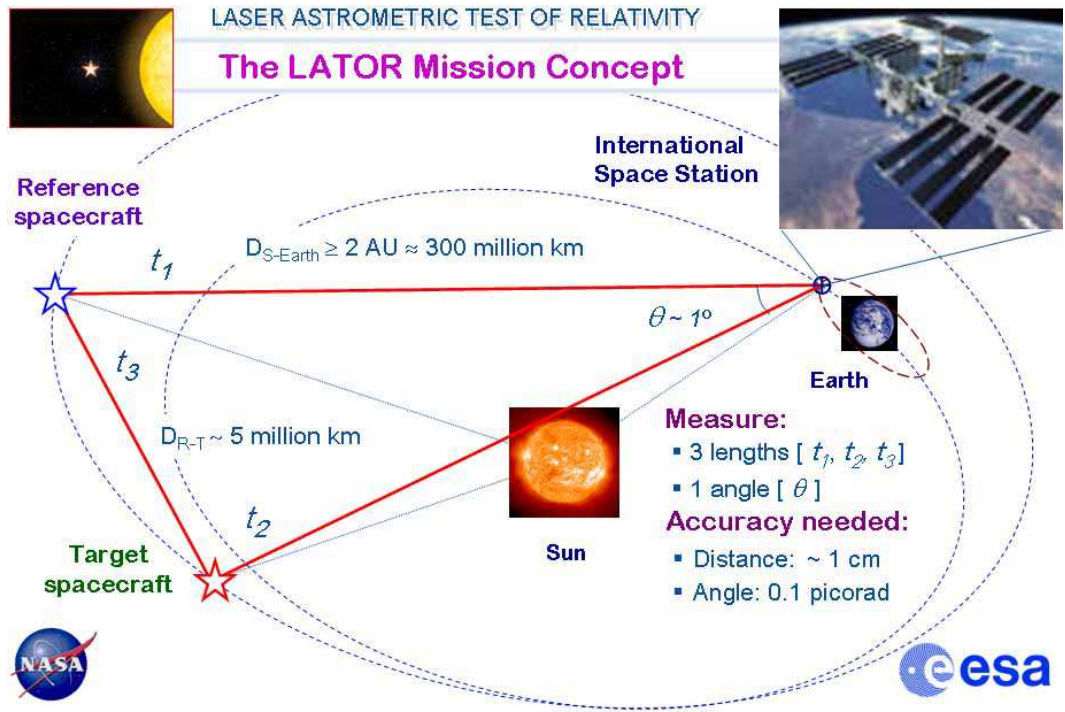,width=140mm}
\end{center}
\vskip -10pt 
  \caption{The overall geometry of the LATOR experiment.  
 \label{fig:lator}}
\vskip -0pt 
\end{figure*} 


When the light deflection in solar gravity is concerned, the magnitude of the first order light deflection effect, as predicted by GR, for the light ray just grazing the limb of the Sun is $\sim1.75$ arcsecond. (Note that 1 arcsecond $\simeq5~\mu$rad; when convenient, below we will use the units of radians and arcseconds interchangeably.) The effect varies inversely with the impact parameter. The second order term is almost six orders of magnitude smaller resulting in  $\sim11$ microarcseconds ($\mu$as) light deflection effect, and which falls off inversely as the square of the light ray's impact parameter (\cite{epstein_shapiro_1980,richter_matzner_1982,Ken_2PPN_87,lator_cqg_2004}). 

This paper discusses the Laser Astrometric Test of Relativity (LATOR)  mission that is designed to directly address the challenges outlined above with an unprecedented accuracy (\cite{lator_cqg_2004}). LATOR will test the cosmologically motivated theories that explain the small acceleration rate of the Universe (so-called `dark energy') via modification of gravity at very large, horizon or super-horizon distances.  This solar system scale experiment would search for a cosmologically-evolved scalar field that is predicted by modern theories of quantum gravity and cosmology, and also by superstring and brane-world models (\cite{dvali}). The physics of a scalar field in the solar system have also been invoked (\cite{BP-2}) as a possible solution to the Pioneer anomaly (\cite{Dittus_etal_05}). The value of the Eddington parameter $\gamma$ may hold the key to the solution of the most fundamental questions concerning the evolution of the universe.  In the low energy approximation suitable for the solar system, modern theories discussed above predict measurable contributions to the parameter $\gamma$ at the level of $(1-\gamma)\sim 10^{-6}-10^{-8}$; detecting this deviation is LATOR's primary objective.  With the accuracy of $1\times10^{-9}$, this mission could discover a violation or extension of general relativity, and/or reveal the presence of any additional long range interaction. 

The paper is organized as follows:  Section \ref{sec:lator} provides the overview for the LATOR experiment, including the preliminary mission design. In Section \ref{sec:lator_current} we discuss the current optical design for the LATOR flight system. We also present the expected performance for the LATOR instrument. Section \ref{sec:conc} discusses the next steps that will be taken in the development of the LATOR mission.  

\section{The LATOR Mission}
\label{sec:lator}

The LATOR mission architecture uses an evolving light triangle formed by laser ranging between two spacecraft (placed in $\sim$1 AU heliocentric orbits) and a laser transceiver terminal on the International Space Station (ISS), via European collaboration.  The objective is to measure the gravitational deflection of laser light as it passes in extreme proximity to the Sun (see Figure \ref{fig:lator}).  To that extent, the long-baseline ($\sim$100 m) fiber-coupled optical interferometer on the ISS will perform differential astrometric measurements of the laser light sources on the two spacecraft as their lines-of-sight pass behind the Sun.  As seen from the Earth, the two spacecraft will be separated by about 1$^\circ$, which will be accomplished by a small maneuver immediately after their launch (\cite{lator_cqg_2004,stanford_texas}). This separation would permit differential astrometric observations to an accuracy of $\sim 10^{-13}$ radians needed to significantly improve measurements of  gravitational deflection of light by the solar gravity.

\begin{figure*}[t!]
 \begin{center}
\noindent    
\psfig{figure=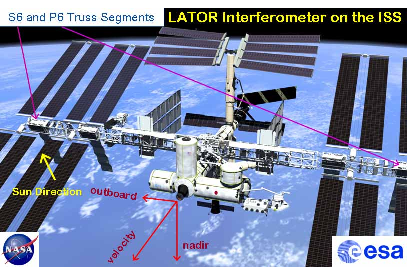,width=115mm}
%
\psfig{figure=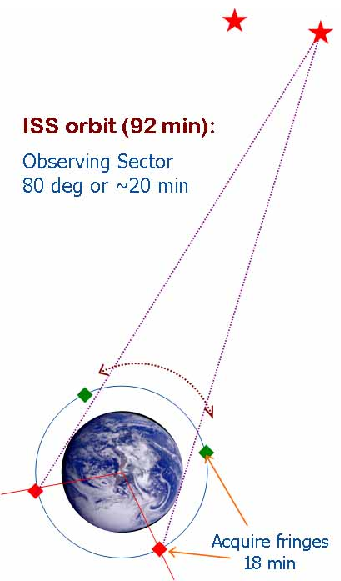,width=4.4cm}
\end{center}
\vskip -10pt 
  \caption{Left: Location of the LATOR interferometer on the ISS. To utilize the inherent ISS sun-tracking capability, the LATOR optical packages will be located on the outboard truss segments P6 and S6 outwards. Right: Signal acquisition for each orbit of the ISS; variable baseline allows for solving fringe ambiguity.  
 \label{fig:lator_sep_angle}}
%
\vskip 5pt 
\hskip -10pt 
\begin{minipage}[b]{.46\linewidth}
\centering \psfig{figure=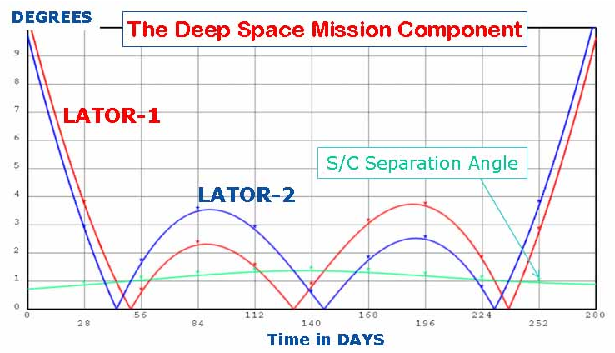,width=118mm}
\end{minipage}
\hfill  
\begin{minipage}[b]{.34\linewidth}
\centering 
\vbox{\psfig{figure=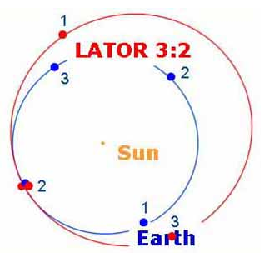,width=5.7cm}\\\vskip26pt}
\end{minipage}
\caption{Left: The Sun-Earth-Probe angle during the period of 3 occultations (two periodic curves) and the angular separation of the spacecraft as seen from the Earth (lower smooth line). Time shown is days from the moment when one of the spacecraft are at 10º distance from the Sun. Right: View from the North Ecliptic of the LATOR spacecraft in a 3:2 resonance. The epoch is taken near the first occultation.  
 \label{fig:lator_sep_angle2}}
\vskip -5pt 
\end{figure*}


To enable the primary objective, LATOR will place two spacecraft into a heliocentric orbit, so that observations may be made when the spacecraft are behind the Sun as viewed from the ISS (see Figures~\ref{fig:lator_sep_angle},\ref{fig:lator_sep_angle2}).  To avoid having to make absolute measurements, the spacecraft will be placed in a 3:2 Earth resonant orbit that provides three observing sessions during the initial 21 months after the launch, with the first session starting in 15 months (\cite{lator_cqg_2004}). Such an orbit provides significant variation of the impact parameter (i.e. distance between the beam and the center of the Sun): the parameter will vary from 10 to 1 solar radii over a period of $\sim$20 days. The three arms of the triangle will be monitored with laser ranging, based on the time-of-flight measurements accurate to $\sim 1$ cm. From three measurements one calculates the Euclidean value for any angle in this triangle.  

As is evident from Figure \ref{fig:lator}, the key element of the LATOR experiment is a redundant geometry optical truss to measure departure from Euclidean geometry ($\sim 8\times 10^{-6}$) caused by the solar gravity field.  This departure is shown as a difference between the calculated Euclidean value for an angle in the triangle and its value directly measured by the interferometer.  This discrepancy, which results from the curvature of the space-time around the Sun and can be computed for every alternative theory of gravity, constitutes LATOR's signal of interest.  The built-in redundancy eliminates the need for drag-free spacecraft for high-accuracy navigation (\cite{lator_cqg_2004}). 
The uniqueness of this mission comes with its built-in geometrically redundant architecture that enables LATOR to measure the departure from Euclidean geometry to a very high accuracy. The precise  measurement of this departure constitutes the primary mission objective.

\subsection{Science with LATOR}

LATOR is a Michelson-Morley-type experiment designed to test the pure tensor metric nature of gravitation - a fundamental postulate of Einstein's theory of general relativity (\cite{lator_cqg_2004}).  With its focus on gravity's action on light propagation it complements other tests which rely on the gravitational dynamics of bodies.  By using a combination of independent time-series of highly accurate gravitational deflection of light in the immediate proximity to the Sun along with measurements of the Shapiro time delay on the interplanetary scales (to a precision respectively better than $10^{-13}$~rad and 1 cm), LATOR will significantly improve tests of relativistic gravity.  

The primary mission objective is to measure the key post-Newtonian Eddington parameter $\gamma$ with an accuracy of a part in 10$^9$.  This parameter, whose value in GR is unity, is perhaps the most fundamental PPN parameter, in that $(1-\gamma)$ is a direct measure of the presence of a new interaction in gravitational theory (\cite{Damour_Nordtvedt_1993a,Damour_EFarese96,lator_cqg_2004}). Within perturbation theory for such theories, all other PPN parameters to all relativistic orders collapse to their GR values in proportion to $(1-\gamma)$. This is why the measurement of the first order light deflection effect at a level of accuracy comparable with the second-order contribution would provide the crucial information separating alternative scalar-tensor theories of gravity from the general theory of relativity (\cite{Ken_2PPN_87}) and also to probe possible scenarios for the quantization of gravity, as well as testing modern theories of cosmological evolution (\cite{Damour_Nordtvedt_1993a,DamourPolyakov94,dvali,DPV02})  discussed in the previous section.  LATOR is designed to directly address this issue with an unprecedented accuracy and in its search, LATOR goes a factor 30,000 beyond the present best result, Cassini's 2003 test (\cite{cassini_ber}). It will also reach ability to measure the next, i.e. post-post-Newtonian, order ($\propto G^2$) of light deflection with accuracy to 1 part in $10^3$. (Note that it has been shown that there are no new higher-order PPN parameters entering the ${\cal O}(G^2)$ order of light deflection (\cite{Damour_EFarese96}).) Other mission objectives are presented in Table~\ref{tab:summ_science}. LATOR's primary measurement pushes to unprecedented accuracy the search for cosmologically relevant scalar-tensor theories of gravity by looking for a remnant scalar field in today's solar system.

\def\reff{\vskip 4pt \par \hangindent 12pt \noindent}
\begin{table}[t!]
\caption{LATOR Mission Summary: Science Objectives 
\label{tab:summ_science}}
\vskip 5pt
\begin{tabular}{m{8.2cm}} \hline \hline\\[-8pt] 

{\it Qualitative Objectives: }

\reff $\bullet$\hskip6pt
To test the metric nature of the Einstein's general theory of relativity in the most intense gravitational environment available in the solar system -- the extreme proximity to the Sun

\reff $\bullet$\hskip6pt
To test alternative theories of gravity and cosmology, notably scalar-tensor theories, by searching for cosmological remnants of scalar field in the solar system

\reff $\bullet$\hskip6pt
To verify the models of light propagation and motion of the gravitationally-bounded systems at the second post-Newtonian order (i.e. including effects $\propto G^2$)\\[4pt] 

{\it Quantitative Objectives: }

\reff $\bullet$\hskip6pt
To measure the key Eddington PPN parameter $\gamma$ with accuracy of 1 part in 10$^{9}$ -- a factor of 30,000 improvement in the tests of gravitational deflection of light

\reff $\bullet$\hskip6pt
To provide direct and independent measurement of the Eddington PPN parameter $\beta$ via gravity effect on light to $\sim0.01$\% accuracy

\reff $\bullet$\hskip6pt
To measure effect of the 2-nd order gravitational deflection of light with accuracy of $\sim1\times 10^{-4}$, including first ever measurement of the post-PPN parameter $\delta$ 

\reff $\bullet$\hskip6pt
To measure the solar quadrupole moment $J_2$ (using the theoretical value of $J_2 \simeq 10^{-7}$) to 1 part in 200, currently unavailable

\reff $\bullet$\hskip6pt
To directly measure the frame dragging effect on light (first such observation and also first direct measurement of solar spin) with $\sim 1\times10^{-3}$ accuracy. 
\\[2pt]\hline 
\end{tabular}
\vskip-10pt
\end{table}
 

The goal of measuring deflection of light in solar gravity with accuracy of one part in $10^{9}$ requires serious consideration of systematic errors. This work requires a significant effort to properly identify the entire set of factors that may influence the accuracy at this level. Fortunately, we initiated this process aided by previous experience in the development of a number of instruments that require similar technology and a comparable level of accuracy (\cite{lator_cqg_2004}). This experience comes with understanding various constituents of the error budget, expertise in developing appropriate instrument models; it is also supported by the extensive verification of the expected  performance with instrumental test-beds and existing flight hardware. Details of the LATOR error budget are being developed and will be published elsewhere, when fully analyzed. Recent covariance studies confirmed the expected mission performance and emphasized the significant potential of the mission (\cite{Ken_lator05,hellings_2005}). 

We shall now consider the LATOR optical design. 

\section{Optical Design}
\label{sec:lator_current}

\begin{figure*}[t!]
 \begin{center}
\noindent    
\psfig{figure=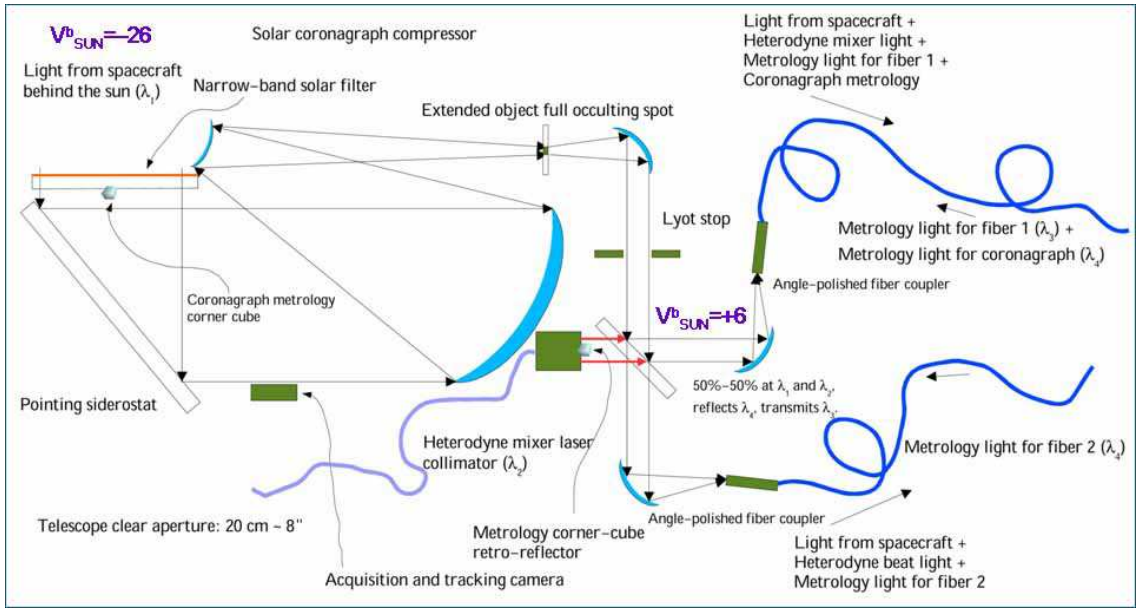,width=150mm}
\end{center}
\vskip -10pt 
  \caption{Basic elements of optical design for the LATOR interferometer: The laser light (together with the solar background) is going through a full aperture ($\sim20$cm) narrow band-pass filter with $\sim 10^{-4}$ suppression properties. The remaining light illuminates the baseline metrology corner cube and falls onto a steering flat mirror where it is reflected to an off-axis telescope with no central obscuration (needed for metrology). It is then enters the solar coronograph compressor by first going through a 1/2 plane focal plane occulter and then coming to a Lyot stop. At the Lyot stop, the background solar light is reduced by a factor of $10^{6}$. The combination of a narrow band-pass filter and coronograph enables the solar luminosity reduction from $V=-26$ to $V=4$ (as measured at the ISS), thus, enabling the LATOR precision observations.
\label{fig:optical_design}}
\vskip -5pt 
\end{figure*}

A single aperture of the interferometer on the ISS consists of three 20 cm diameter telescopes (see Figure \ref{fig:optical_design} for a conceptual design). One of the telescopes with a very narrow bandwidth laser line filter in front and with an InGAs camera at its focal plane, sensitive to the 1064 nm laser light, serves as the acquisition telescope to locate the spacecraft near the Sun.

The second telescope emits the directing beacon to the spacecraft. Both spacecraft are served out of one telescope by a pair of piezo controlled mirrors placed on the focal plane. The properly collimated laser light ($\sim$10W) is injected into the telescope focal plane and deflected in the right direction by the piezo-actuated mirrors. 

The third telescope is the laser light tracking interferometer input aperture which can track both spacecraft at the same time. To eliminate beam walk on the critical elements of this telescope, two piezo-electric X-Y-Z stages are used to move two single-mode fiber tips on a spherical surface while maintaining focus and beam position on the fibers and other optics. Dithering at a few Hz is used to make the alignment to the fibers and the subsequent tracking of the two spacecraft completely automatic. The interferometric tracking telescopes are coupled together by a network of single-mode fibers whose relative length changes are measured internally by a heterodyne metrology system to an accuracy of less than 10 pm.

The spacecraft  are identical in construction and contain a relatively high powered (1 W), stable (2 MHz per hour $\sim$500 Hz per second), small cavity fiber-amplified laser at 1064 nm. Three quarters of the power of this laser is pointed to the Earth through a 10 cm aperture telescope and its phase is tracked by the interferometer. With the available power and the beam divergence, there are enough photons to track the slowly drifting phase of the laser light. The remaining part of the laser power is diverted to another telescope, which points towards the other spacecraft. In addition to the two transmitting telescopes, each spacecraft has two receiving telescopes.  The receiving telescope, which points towards the area near the Sun, has laser line filters and a simple knife-edge coronagraph to suppress the Sun light to 1 part in $10^4$ of the light level of the light received from the space station. The receiving telescope that points to the other spacecraft is free of the Sun light filter and the coronagraph.

In addition to the four telescopes they carry, the spacecraft also carry a tiny (2.5 cm) telescope with a CCD camera. This telescope is used to initially point the spacecraft directly towards the Sun so that their signal may be seen at the space station. One more of these small telescopes may also be installed at right angles to the first one to determine the spacecraft attitude using known bright stars. The receiving telescope looking towards the other spacecraft may be used for this purpose part of the time, reducing hardware complexity. Star trackers with this construction have been demonstrated many years ago and they are readily available. A small RF transponder with an omni-directional antenna is also included in the instrument package to track the craft while they are on their way to assume the orbital position needed for the experiment. 

In the next Section we present elements for the LATOR optical receiver system.  While we focus on the optics for the two spacecraft, the interferometer's architecture includes essentially similar optical components. 

\begin{figure*}[t!]
 \begin{center}
\noindent    
\psfig{figure=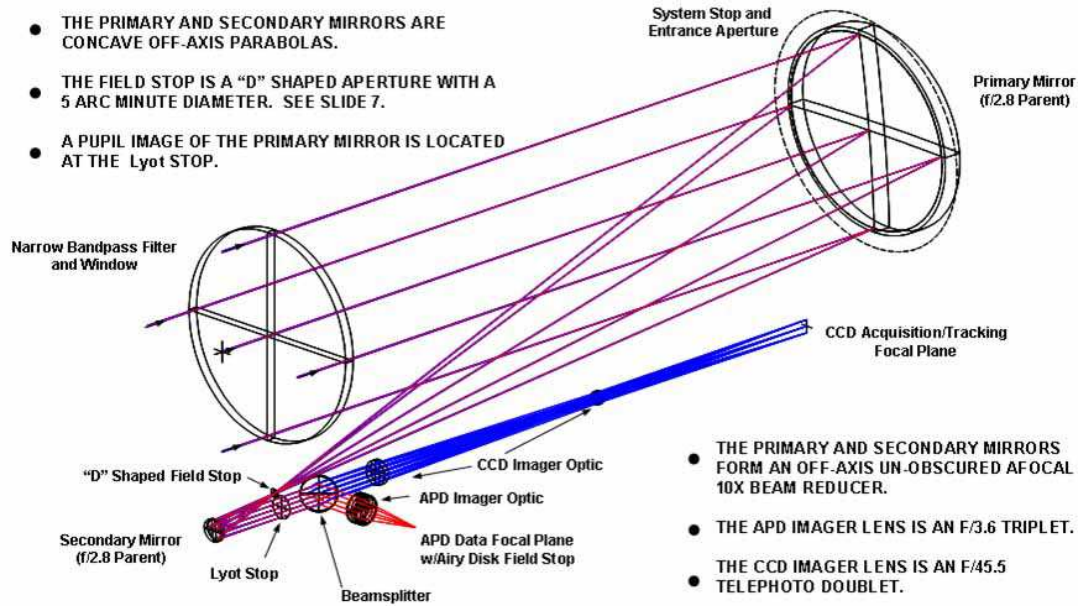,width=146mm}
\end{center}
\vskip -10pt 
  \caption{LATOR receiver optical system layout.  
 \label{fig:lator_receiver}}
\end{figure*} 
\begin{figure*}[t!]
 \begin{center}
\noindent    
\psfig{figure=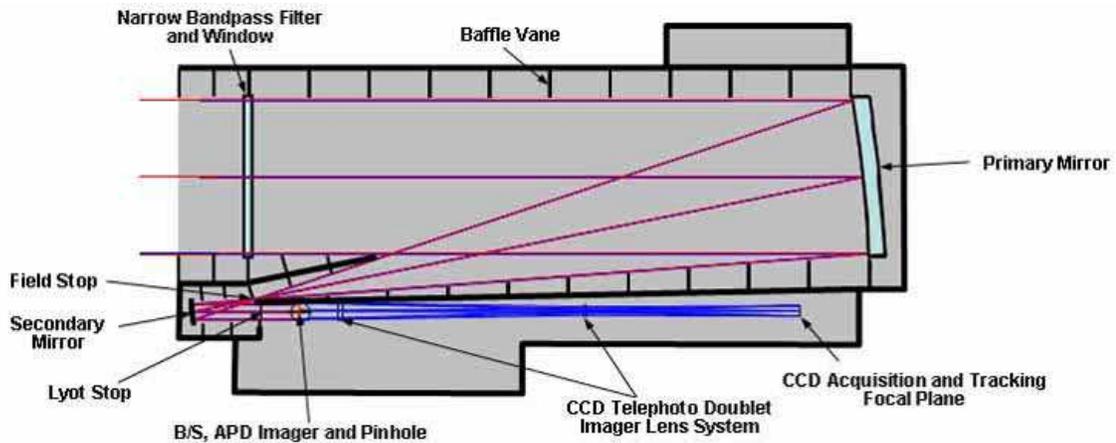,width=150mm}
\end{center}
\vskip -10pt 
  \caption{The LATOR preliminary baffle design.  
 \label{fig:lator_buffle}}
\end{figure*} 

\subsection{The LATOR Optical Receiver System}

The LATOR 100 mm receiver optical system is a part of a proposed experiment. This system is located at each of two separate spacecraft placed on heliocentric orbits, as shown in Figure \ref{fig:lator}. The receiver optical system captures optical communication signals form a transmitter on the ISS, which orbits the Earth. To support the primary mission objective, this system must be able to receive the optical communication signal from the uplink system at the ISS that passes through the solar corona at the immediate proximity of the solar limb (at a distance of no more than 5 Airy disks). Our recent analysis of the LATOR 100 mm receiver optical system successfully satisfied all the configuration and performance requirements (\cite{stanford,hellings_2005}). We have also performed a conceptual design (see Figure \ref{fig:lator_receiver}), which was validated with a ray-trace analysis. The ray-trace performance of the designed instrument is diffraction limited in both the APD and CCD channels over the specified field of view at 1064 nm. The design incorporated the required field stop and Layot stop. A preliminary baffle design has been developed for controlling the stray light.

\subsection{Preliminary Baffle Design}

\begin{figure}[t!]
 \begin{center}
\noindent    
\psfig{figure=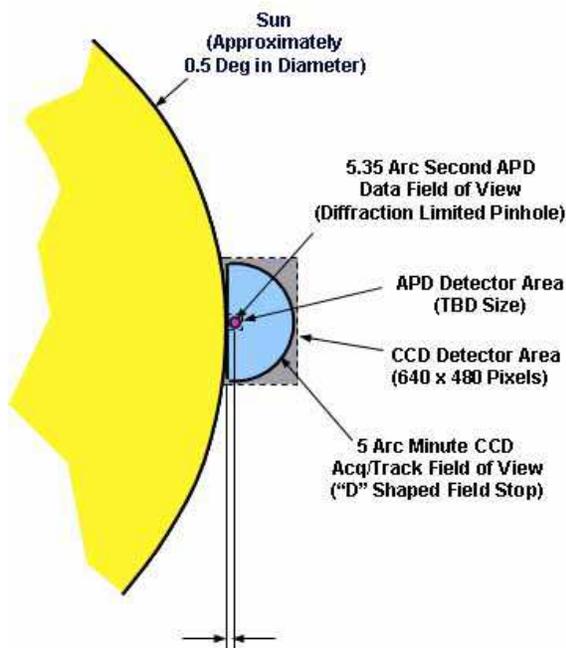,width=77mm}
\end{center}
\vskip -10pt 
  \caption{LATOR focal plane mapping (not to scale).  
 \label{fig:lator_focal}}
\vskip -10pt 
\end{figure}

Figure \ref{fig:lator_buffle} shows the LATOR preliminary baffle design. The out-of-field solar radiation falls on the narrow band pass filter and primary mirror; the scattering from these optical surfaces puts some solar radiation into the FOV of the two focal planes. This imposes some requirements on the instrument design.  
Thus, the narrow band pass filter and primary mirror optical surfaces must be optically smooth to minimize narrow angle scattering. This may be difficult for the relatively steep parabolic aspheric primary mirror surface. However, the field stop will eliminate direct out-of-field solar radiation at the two focal planes, but it will not eliminate narrow angle scattering for the filter and primary mirror.  Finally, the Lyot stop will eliminate out-of-field diffracted solar radiation at the two focal planes. Additional baffle vanes may be needed at several places in the optical system. This design will be further investigated in series of trade-off studies by also focusing on the issue of stray light analysis. Figure \ref{fig:lator_focal} shows the design of the focal plane capping. The straight edge of the `D'-shaped CCD field stop is tangent to the limb of the Sun and it is also tangent to the edge of APD field stop. There is a 2.68 arcsecond offset between the straight edge and the concentric point for the circular edge of the CCD field stop.  The results of the analysis of APD and CCD channels point spread functions can be found in (\cite{stanford}). 

\section{Conclusions}
\label{sec:conc}

The LATOR experiment benefits from a number of advantages over techniques that use radio waves to study the light propagation in the solar vicinity.  The use of monochromatic light enables the observation of the spacecraft almost at the limb of the Sun, as seen from the ISS.  The use of narrowband filters, coronagraph optics, and heterodyne detection will suppress background light to a level where the solar background is no longer the dominant noise source.  The short wavelength allows much more efficient links with smaller apertures, thereby eliminating the need for a deployable antenna.  Advances in optical communications technology allow low bandwidth telecommunications with the LATOR spacecraft without having to deploy high gain radio antennae needed to communicate through the solar corona.  Finally, the use of the ISS not only makes the test affordable, but also allows conducting the experiment above the Earth's atmosphere, the major source of astrometric noise for any ground based interferometer.  This fact justifies the placement of LATOR's interferometer node in space. 

The concept is technologically sound; the required technologies have been demonstrated as part of the international laser ranging activities and optical interferometry programs at JPL. LATOR does not need a drag-free system, but uses a geometric redundant optical truss to achieve a very precise determination of the interplanetary distances between the two micro-spacecraft and a beacon station on the ISS. The experiment takes advantage of the existing space-qualified optical technologies, leading to an outstanding performance in a reasonable mission development time. In addition, the issues of the extended structure vibrations on the ISS, interferometric fringe ambiguity, and signal acquisition on the solar backgrounds have all been analyzed, and do not compromise mission goals.  The ISS is the default location for the interferometer, however, ground- and free-flying versions have also been studied.  While offering programmatic benefits, these options differ in cost, reliability and performance. The  availability of the ISS (via European collaboration) makes this mission concept realizable in the very near future. A recent JPL Team X study confirmed the feasibility of LATOR as a NASA Medium Explorer (MIDEX) class mission. 

LATOR is envisaged as a partnership between NASA and ESA wherein both partners are essentially equal contributors, while focusing on different mission elements: NASA provides the deep space mission components and interferometer design, while building and servicing infrastructure on the ISS is an ESA contribution. The NASA focus is on mission management, system engineering, software management, integration (both of the payload and the mission), the launch vehicle for the deep space component, and operations. The European focus is on interferometer components, the initial payload integration, optical assemblies and testing of the optics in a realistic ISS environment. The proposed arrangement would provide clean interfaces between familiar mission elements.

This mission may become a 21st century version of the Michelson-Morley experiment in the search for a  cosmologically evolved scalar field in the solar system. As such, LATOR will lead to very robust advances in the tests of fundamental physics: it could discover a violation or extension of general relativity, and/or reveal the presence of an additional long range interaction in the physical law.  There are no analogs to the LATOR experiment; it is unique and is a natural culmination of solar system gravity experiments.

\begin{acknowledgements}

The work described here was carried out at the Jet Propulsion Laboratory, California Institute of Technology, under a contract with the National Aeronautics \& Space Administration.
\end{acknowledgements}

\end{document}